\documentclass[12pt]{article}
\usepackage{amssymb,amsmath,slashed,latexsym}
\usepackage{float}
\allowdisplaybreaks
\newcommand{\ft}[2]{{\textstyle\frac{#1}{#2}}}

\def\rme{{\rm e}}
\def\rmi{{\rm i}}

\newcommand{\bbox}{\lower.2ex\hbox{$\Box$}}
\newsavebox{\uuunit}
\sbox{\uuunit}
    {\setlength{\unitlength}{0.825em}
     \begin{picture}(0.6,0.7)
        \thinlines
        \put(0,0){\line(1,0){0.5}}
        \put(0.15,0){\line(0,1){0.7}}
        \put(0.35,0){\line(0,1){0.8}}
       \multiput(0.3,0.8)(-0.04,-0.02){12}{\rule{0.5pt}{0.5pt}}
     \end {picture}}

\csname @addtoreset\endcsname{equation}{section}

   \usepackage[nosort]{cite}
   \pdfoutput=1
  \usepackage[pdftex]{hyperref}
\newcommand{\dr}{\raise.3ex\hbox{$\stackrel{\leftarrow}{\delta  }$}{}}
\newcommand{\dl}{\raise.3ex\hbox{$\stackrel{\rightarrow}{\delta }$}{} }
\newcommand{\pl}{\raise.3ex\hbox{$\stackrel{\rightarrow}{\partial }$}{} }

\begin{document}

\begin{titlepage}
\begin{flushright}
CERN-TH-2019-225
\end{flushright}
\vspace{.5cm}
\begin{center}
\baselineskip=16pt
{\LARGE  Supercurrents in N=1 Minimal Supergravity in the Superconformal Formalism  
}\\
\vfill
{\large  {\bf Sergio Ferrara}$^{1,2}$, {\bf Marine Samsonyan}$^3$, \\[2mm] {\bf Magnus Tournoy}$^4$ and {\bf Antoine Van Proeyen$^4$}, } \\
\vfill

{\small$^1$ Theoretical Physics Department, CERN CH-1211 Geneva 23, Switzerland\\\smallskip
$^2$ INFN - Laboratori Nazionali di Frascati Via Enrico Fermi 40, I-00044 Frascati, Italy\\\smallskip
$^3$ A. I.  Alikhanyan National Laboratory (Yerevan Physics Institute), 2 Alikhanian Brothers St., 0036 Yerevan, Armenia\\\smallskip
$^4$   KU Leuven, Institute for Theoretical Physics, Celestijnenlaan 200D, B-3001 Leuven,
Belgium  \\[2mm] }
\end{center}
\vfill
\begin{center}
{\bf Abstract}
\end{center}
{\small
We discuss the Einstein tensor, the supercurrent and their conservation laws of old and new minimal formulations of supergravity in the superconformal approach. The variation of the action with respect to the gauge field of the $R$-symmetry in the conformal approach (the auxiliary field in the super-Poincar\'{e} action) allows to find  the Einstein tensor and supercurrent in any curved background. Hence generalized expressions for their Ward identities follow. This proceeding is based on \cite{Ferrara:2017yhz,Ferrara:2018dyt}.
} \vfill

\hrule width 3.cm
{\footnotesize \noindent e-mails: Sergio.Ferrara@cern.ch, marine.samsonyan@cern.ch, magnus.tournoy@kuleuven.be,\\
antoine.vanproeyen@fys.kuleuven.be }
\end{titlepage}

\addtocounter{page}{1}
 \tableofcontents{}
\newpage
      
\section{Introduction}
In general for every rigid symmetry there is a conserved current upon using equations of motion. Currents can be found from gauge couplings.
 For example, from the Lagrangian of electromagnetic field $S=\int d^4x [-\frac{1}{4g^2} F_{\mu\nu}F^{\mu\nu} + A^\mu J_\mu +...]$, the variation with respect to $A_\mu$ gives a current $\partial^\mu F_{\mu\nu}=J_\nu$, which is conserved upon using the field equations of $A_\mu$, $\partial^\mu J_\mu \approx  0$ \footnote{We will indicate equations valid modulo equations of motion by $\approx $.}. Likewise, for pure gravity one finds  $G_{\mu\nu}\approx 0$.
The canonical energy momentum-tensor is symmetric due to Lorentz rotation invariance: $T_{\mu\nu}=T_{\nu\mu}$. It is also conserved once the field equations of the theory are used. Callan, Coleman and Jackiw (CCJ) have proved in \cite{Callan:1970ze} that the energy-momentum tensor can be made traceless once the theory possesses a scale and conformal invariance.
\begin{eqnarray}
\Theta_{\mu\nu}=T_{\mu\nu}+\text{improvement term}\, ,\quad
\partial^\mu\Theta_{\mu\nu}=0, \quad \Theta_{\mu\nu}=\Theta_{\nu\mu},\quad \Theta_\mu^{\,\,\,\mu}=0\, .
\end{eqnarray}
This improved energy-momentum tensor found its place in the supercurrent multiplet $J_{\alpha\dot{\alpha}}$. In rigid supersymmetry the conservation law of a supercurrent $J_{\alpha \dot \alpha }$ expressed through its supercurrent divergence is\cite{Ferrara:1975pz,Ferrara:1977mv,Gates:1983nr,Komargodski:2009pc,Komargodski:2010rb,Kuzenko:2010ni}
\begin{equation}
  \overline{D}^{\dot \alpha }J_{\alpha \dot \alpha }\approx  D_\alpha  Y +\omega _\alpha \,,
 \label{conservationEqn}
\end{equation}
where $Y$, $\omega _\alpha $ are chiral superfields and moreover $\omega _\alpha $ satisfies a reality condition $D_\alpha \omega ^\alpha = \overline{D}_{\dot \alpha }\omega ^{\dot \alpha}$.
In linearized supergravity \cite{Ferrara:1977mv,Komargodski:2010rb} the current couples to the Einstein multiplet $E_{\alpha \dot \alpha }$:
\begin{equation}
  E_{\alpha \dot \alpha }\approx -\kappa ^2 J_{\alpha \dot \alpha }\,,
 \label{E+J0}
\end{equation}
with their Ward identities in old and new minimal formulations:
\begin{eqnarray}
\overline{D}^{\dot \alpha }E_{\alpha \dot \alpha }= & D_\alpha {\cal R},\qquad \overline{D}^{\dot \alpha }E_{\alpha \dot \alpha }= & W_\alpha ^L\,,
\end{eqnarray}
where ${\cal R}$ is a scalar chiral multiplet and $W_\alpha ^L$ is a spinor chiral multiplet. 
The trace equation becomes in the two formulations
\begin{equation}
  {\cal R}\approx -\kappa ^2 Y\,,\qquad W_\alpha ^L\approx -\kappa ^2\omega  _\alpha\,.
 \label{XYW}
\end{equation}
The equations (\ref{E+J0}) and (\ref{XYW}) are the supergravity extensions of the Einstein equation and its trace:
\begin{align}
  G_{\mu \nu }\approx  \kappa ^2 T_{\mu \nu }\,,\qquad
  R \approx -\kappa ^2 T_\mu {}^\mu \,.\label{traceEinstein}
\end{align}

The content of this proceeding is based on \cite{Ferrara:2017yhz,Ferrara:2018dyt}. We will give fully non-linear expressions for the Ward identities of supercurrents using the superconformal techniques explained in \cite{Freedman:2012zz}.

\section{The Improved Energy-Momentum Tensor}
Let us shortly review the CCJ improved tensor by considering a Lagrangian of a single scalar field \footnote{For a discussion of more general settings see \cite{Ferrara:2017sba,Ferrara:2018ucv}.}
\begin{equation}
S=\int d^4 x \sqrt{-g}(\ft 12 \partial _\mu \varphi \partial ^\mu \varphi - V(\varphi)).
\end{equation}
It has a symmetric and conserved, but not traceless canonical energy-momentum tensor.
The improved energy-momentum tensor is defined with an improvement term
\begin{eqnarray}
\Theta_{\mu\nu} =T_{\mu\nu}-\frac{1}{6}(\partial_\mu\partial_\nu-g_{\mu\nu}\Box) \varphi^2 ,
\end{eqnarray}
which is obviously conserved $\partial ^\mu \Theta_{\mu\nu} =0$. The second term now contributes to the trace and gives a traceless improved tensor upon using $\varphi$ field equation provided the potential is quartic. The latter ensures a conformal invariance
\begin{eqnarray}
\Theta_\mu{}^\mu \approx 0\, .
\end{eqnarray}
The improved energy-momentum tensor can also be obtained from the conformal invariant gravity-coupled action.
For a conventional gravity theory
\begin{eqnarray}
S=\int d^4 x \sqrt{-g}\left[ \ft{1}{2}\kappa^{-2}R +{\cal L}_M\right] \, ,
\end{eqnarray}
the variation with respect to the metric gives
\begin{eqnarray}
G_{\mu\nu}\approx\kappa^2 T_{\mu\nu}\, .
\end{eqnarray}
To have conformal and Weyl symmetry, we should replace $\kappa^{-2}  \to \kappa^{-2}-\tfrac{1}{6}\varphi ^2 $. After doing this replacement and considering the single scalar field for ${\cal L}_M$, we get
\begin{eqnarray}
{\cal S}=\int d^4x \sqrt{-g}\left[\ft12(\kappa^{-2}-\tfrac{1}{6} \varphi^2)R+\ft12 g^{\mu \nu }\partial _\mu \varphi\partial _\nu \varphi -V(\varphi) \right] \, ,\\
G_{\mu \nu }\approx \kappa^2\Theta_{\mu \nu }^{\rm c}\,,\qquad \Theta_{\mu \nu }^{\rm c}= T_{\mu \nu }-\ft{1}{6}(\nabla _\mu\partial_\nu-g_{\mu\nu}\nabla ^\rho \partial _\rho ) \varphi^2+\tfrac{1}{6} \varphi^2G_{\mu \nu }\,.
\end{eqnarray}
It is easy to check that $\nabla^\mu\Theta_{\mu \nu }^{\rm c}\approx0$, and $\Theta^{\rm c\,\,\, \mu}_\mu \approx 0$ for quartic $V(\varphi)$.
Thus, to recapitulate, we have the following picture
\begin{align}
&\text{Rigid Poincar\'e}& \qquad &T_{\mu\nu}=\partial_\mu \varphi \partial_\nu \varphi +g_{\mu\nu} L\, ,&\\
&\text{Rigid conformal}& \qquad &\Theta_{\mu\nu} =T_{\mu\nu}-\frac{1}{6}(\partial_\mu\partial_\nu-g_{\mu\nu}\Box) \varphi^2\, ,&\\
&\text{Local conformal}& \qquad &\Theta_{\mu \nu }^{\rm c}= T_{\mu \nu }-\ft{1}{6}(\nabla _\mu\partial_\nu-g_{\mu\nu}\nabla ^\rho \partial _\rho ) \varphi^2+\tfrac{1}{6} \varphi^2G_{\mu \nu }\, .
\end{align}

\subsection{From Conformal Action}
These formulations can be obtained from a conformal action, containing apart from the physical field $\varphi $ also a compensating scalar $\varphi _0$. Both have then Weyl weight~1. The action reads
\begin{eqnarray}
 S&=&\int d^4 \sqrt{g}\left[-\ft12\varphi _0\Box^C \varphi_0+\ft12\varphi\Box^C\varphi +\lambda \varphi ^4\right]  \nonumber\\
  &=&\int d^4 \sqrt{g}\left[\ft12\partial _\mu \varphi_0\partial^\mu  \varphi_0 - \ft12\partial _\mu \varphi\partial^\mu  \varphi+\ft1{12}(\varphi _0^2-\varphi ^2) R  +\lambda \varphi ^4\right]\,.
\end{eqnarray}
To gauge fix the compensating scalar, one should choose a frame. The Einstein frame corresponds to fixing $\varphi _0^2=\varphi ^2 + 6 \kappa ^{-2}$. In this case one observes that no conformal part in the action is present, and hence there is no traceless energy-momentum tensor. Thus another frame, called a conformal frame, should be chosen. It corresponds to taking $\varphi _0^2= 6 \kappa ^{-2}$,  which provides conformal action and traceless energy-momentum tensor.

\section{Supercurrent and Einstein Tensor from Superconformal Approach}
The actions of chiral multiplets in the superconformal setup are symbolically obtained from
\begin{equation}
  S=\left[N(X^I,\bar X^{\bar I})\right]_D + \left[{\cal W}(X^I)\right]_F\, .
\end{equation}
 $X^I$ are chiral multiplets with $(1,1)$ Weyl and chiral weights. $I=0,...,n$, where $n$ is the number of physical multiplets, $0$-index is the multiplet that compensates the superconformal symmetry. The D-term is real of Weyl and chiral weights $(2,0)$ and the F-term is holomorphic of weights $(3,3)$. See \cite{Freedman:2012zz} or Appendix A  of \cite{Ferrara:2017yhz} for detailed explanation of the notation. Since pure supergravity is given by  $ N= -3X^0\bar X^{\bar 0}$, we can reorganize the variables $X^I$ in the form
$ S^0 = X^0\,,\quad S^i= \frac{X^i}{X^0}\,,\quad i=1,\ldots ,n\,,$ and write
\begin{equation}
  N(X,\bar X)=S^0\bar S^{\bar 0} \Phi (S,\bar S)\,,\qquad {\cal W}(X)=(S^0)^3W(S)\, .
\end{equation}
Furthermore, to separate pure supergravity from the matter part, we can make the splitting $\Phi(S,\bar S)=-3+ 3\Phi_M (S, \bar S)$ such that 
\begin{equation}
N(X,\bar X)=S^0\bar S^{\bar 0}(-3 + 3\Phi _{\rm M}(S,\bar S))=N^G+N^M, \qquad {\cal W}(X)= (S^0)^3 W(S)\, ,\label{splaction}
\end{equation}
where $N^G$ denotes the pure supergravity and $N^M$ is the matter part.
The field equation for the compensating multiplet, $I=0$ can be written as
\begin{equation}
  {\cal R}+\frac{Y}{(S^0)^2}\approx 0 \quad \text{with} \quad Y\equiv - 2(S^0)^3 \Delta W +S^0 T(\bar S^{\bar 0} \Delta K)\, .\label{RY}
\end{equation}
The scalar curvature is defined in terms of the compensating multiplet as ${\cal R}\equiv  \frac{1}{S^0}T(\bar S^{\bar 0}) $  \cite{Kugo:1983mv}, where the operation $T$ is the superconformal version of the superspace operation $\bar D^2$. In the conformal frame $S^0= \kappa^{-1}$, we have $ {\cal R}+\kappa^2 Y\approx 0$.  Hence ${\cal R}\approx 0$ for the conformal case. The equation \eqref{RY} is related to the global formulae in
\cite{Clark:1995bg,Komargodski:2009rz, Komargodski:2010rb,Kuzenko:2010am,Kuzenko:2010ni,Korovin:2016tsq}. In \eqref{RY} we defined
\begin{align}
 \Delta K &\equiv -\frac{1}{3\bar S^{\bar 0}}\left(N_0+3\bar S^{\bar 0}\right)=  S^i\Phi_{{\rm M}\,i}-\Phi_{\rm M}\, ,\\
 \Delta W&\equiv \frac{1}{3(S^0)^2}{\cal W}_0= W-\frac{1}{3}S^iW_i\,. \nonumber
\end{align}
The conformal case corresponds to $\Delta K=\Delta W=0$.  Thus $W$ is homogeneous of rank $3$ and $\Phi_M$ is homogeneous of rank 1 both in $S^i$ and $\bar S^i$, $\Phi_{Mi\bar j}$ has degree zero. One field $\Phi_M= S\bar S$ corresponds to the conformally coupled scalar of CCJ.

\section{Non-linear Form of Ferrara-Zumino Equations of 1975}
We would like to generalize the FZ equation $\overline{D}{}^{\dot \alpha }E_{\alpha \dot \alpha }= D_\alpha {\cal R}$ to the non-linear level. In the superconformal  
formulation nonlinearities come from two sources: compensator dependence and coupling to matter. The non linear version of $E_{\alpha \dot \alpha }$ we denote by 
${\cal E}_{\alpha \dot \alpha}$. Let's add the compensating scalar to some power in the FZ equation:
\begin{eqnarray}
\overline{D}^{\dot{\alpha}} {\cal E}_{\alpha \dot{\alpha}} =(S^0)^{k_1} (\bar{S}^{\bar 0})^{k_2} D_\alpha \left(\frac{{\cal R}}{S^0}\right)\, .
\end{eqnarray}
Matching the Weyl and chiral weights on both sides fixes uniquely the powers of $S^0$ and $\bar{S}^0$ as $k_1+k_2=w, \, k_1-k_2=3$. The $S^0$ under the derivative in the denominator on the right hand side is needed to provide a real ${\cal E}_{\alpha\dot\alpha}$.  Thus the non-linear version of FZ equation reads
\begin{equation}
  \overline{{\cal D}}{}^{\dot \alpha }{\cal E}_{\alpha \dot \alpha }= (S^0)^3 {\cal D}_\alpha \left(\frac{{\cal R}}{S^0}\right)\, ,\label{WardE}
\end{equation}
which is the generalized Bianchi identity. Using ${\cal R}$ in \eqref{RY}, this equation can be solved. Its flat limit reproduces the known formula
\begin{eqnarray}
  E_{\alpha \dot\alpha} = - 4i \bar S^0 \overset{\leftrightarrow}{\partial}_{\alpha \dot\alpha }S^0-2(D_\alpha S^0)(\bar D_{\dot\alpha  }\bar S^{\bar 0})\,.
\end{eqnarray}

\section{The Supercurrent}
Since we have used an explicit splitting in the action \eqref{splaction}, we can use the relation between ${\cal R}$ and $Y$  and ${\cal E}_{\alpha\dot\alpha}+J_{\alpha\dot\alpha}\approx 0$ to find the conservation law for the supercurrent
\begin{equation}
\overline{{\cal D}}^{\dot\alpha}  J_{\alpha\dot\alpha}\approx (S^0)^3 {\cal D}_\alpha \left(\frac{Y}{(X^0)^3}\right)
\approx - (S^0)^3 {\cal D}_\alpha \left(2\Delta W -(S^0)^{-2}T\left(\bar S^{\bar0}\Delta K\right)\right)\label{WardJ}\,.
\end{equation}
One observes that $\overline{\cal{D}}^{\dot\alpha} J_{\alpha\dot\alpha}\approx 0$ in the conformal case.
 ${\cal E}$ and ${\cal J}$ that satisfy \eqref{WardE} and \eqref{WardJ} can be written in components as
\begin{eqnarray}
  &J_\mu&  =-\Phi _M{\cal E}_\mu  +2i \bar X^{\bar 0}\Phi _{M\,\bar i}\bar \chi ^{\bar 0}\gamma _\mu \chi ^{\bar i}
  +2i X^0\Phi _{M\,i}\bar \chi ^i\gamma _\mu \chi ^{0}\nonumber\\
&&  + 2i X^0\bar X^{\bar 0}\left[2(\Phi _{M\,i}{\cal D}_\mu S^i-\Phi _{M\,\bar i}{\cal D}_\mu \bar S^{\bar i})-\Phi _{M\,i\bar j}\bar\chi ^i\gamma _\mu \chi ^{\bar j} \right]\,,\\
&{\cal E}_\mu &= 4i X^0{\cal D}_\mu \bar X^{\bar 0}-4i \bar X^{\bar 0}{\cal D}_\mu X^0+2i \overline{\chi }^0\gamma _\mu \chi ^{\bar 0}\,.
\end{eqnarray}
In the conformal frame $ X^0=\kappa^{-1}$ and  $\Omega^0=0 $, ${\cal E}_\mu $ does not depend on matter fields, and we have
\begin{align}
{\cal E}_\mu = -8\kappa ^{-2} A_\mu\,,
J_\mu =8\kappa ^{-2} \Phi_M A_\mu +2i\kappa ^{-2}\left[2(\Phi _{M\,i}{\cal D}_\mu S^i-\Phi _{M\,\bar i }{\cal D}_\mu \bar S^{\bar i})-\Phi _{M\,i\bar j}\bar\chi ^i\gamma _\mu \chi ^{\bar j} \right],
\end{align}
which is the generalization of the CCJ. 

\subsection{$A_\mu $ Field Equation}
One can observe that 
\begin{equation}
  -\ft34 ({\cal E}_\mu+ J_\mu )= i N_{\bar I}{\cal D}_\mu \bar X^{\bar I}- i N_I{\cal D}_\mu X^I+\ft{1}{2} i N_{I\bar J}\overline{\Omega }^I\gamma _\mu \Omega ^{\bar J}\label{AFE}\,.
\end{equation}
Since $[{\cal W}]_F$ does not  involve  $A_\mu $ (the gauge field of the $R$-symmetry in the conformal approach, and is the auxiliary field in the super-Poincar\'{e} action), \eqref{AFE} gives the $A_\mu $ field equation
\begin{equation}
  e^{-1}\frac{\delta }{\delta A^\mu }[N(X,\bar X)]_D =-\ft34 ({\cal E}_\mu+ J_\mu )\,.
\end{equation}
This expression is a superconformal primary, and can be used as first component of a superconformal and Einstein tensor multiplet. The full multiplet was found in \cite{Ferrara:2017yhz}. In \cite{Ferrara:2018dyt}  it was shown that the variation of the action with respect to the field $A_\mu$ always gives ${\cal E}+{\cal J}$ as far as the action is invariant under all superconformal transformations and the covariantized field equations \cite{Vanhecke:2017chr} are used. In the following we will consider new and old minimal setups with Fayet-Iliopoulos (FI) term and write Ward identities for these theories using techniques of \cite{Ferrara:2018dyt} . 

\section{Old Minimal Supergravity with a FI Term}
Let us consider the old minimal supergravity when FI term is present. The basic Lagrangian is 
\begin{equation}
 {\cal L}= \left[N(X^I,\bar X^I) e^{\xi V}\right]_D + \left[W^\alpha W_\alpha \right]_F\,,\qquad \{X^I\}=\{X^0,\,X^i\}\,,
\end{equation}
where $\xi $ is the (dimensionless) FI constant. $X^0$ transforms under an abelian gauge group gauged by a real multiplet $V$.
\begin{equation}
X^0\rightarrow X^0 e^{-\xi \Lambda}, \quad V\rightarrow V+\Lambda+\bar{\Lambda}, \quad S^i\ \mbox{ inert.}
\end{equation}
The super-Einstein equations are obtained from the Lagrangian by variation of the auxiliary field $A_\mu $
\begin{equation}
  \ft14i \gamma ^\mu _{\alpha \dot \alpha } e^{-1}\frac{ \delta {\cal L}}{\delta A^\mu }=  {\cal E}^{om,V}_{\alpha \dot \alpha }(X^0,V) + J_{\alpha \dot \alpha }(X^0,V,S^i)+{\cal E}^W_{\alpha \dot \alpha }\approx 0\,,\label{feom}
\end{equation}
where ${\cal E}^W_{\alpha \dot \alpha }=4W _{\dot \alpha }W_\alpha$, $W_\alpha \equiv  T {\cal D}_\alpha V= -\ft12\rmi \lambda _\alpha$ with
$\lambda _\alpha $ being the left-projected gaugino in the gauge multiplet. It was proven in \cite{Ferrara:2018dyt} that $ {\cal E}^{om,V}_{\alpha \dot \alpha }$ satisfies the following Ward identity
\begin{equation}
 \bar{{\cal D}}^{\dot \alpha }{\cal E}^{om}_{\alpha \dot \alpha }
   =(e^{\xi V}X^0)^3{\cal D}_\alpha \left[\frac{e^{-3\xi V}T\left(e^{\xi V}\bar X^0\right)}{(X^0)^2}\right]+ 3\xi X^0 e^{\xi V} \bar{X}^0 W_\alpha\,.
\end{equation}
The superspace geometry that encodes this Ward identity was called chirally extended supergravity in \cite{deWit:1978ww}. Note that $\xi \rightarrow 0$ gives the pure old minimal.

From the field equation \eqref{feom} one obtains the $J_{\alpha\dot{\alpha}}$ Ward identity using the splitting \eqref{splaction} and the field equations of auxiliary fields
\begin{equation}
 \overline{{\cal D}}^{\dot \alpha }J_{\alpha \dot \alpha }
   \approx \ft13 (\rme^{\xi V}X^0)^3{\cal D}_\alpha \left[\frac{e^{-3\xi V}T\left(\rme^{\xi V}N_0^{\rm{M}}\right)}{(X^0)^2}\right]-\xi  W_\alpha N^{\rm{M}}\rme^{\xi V}\,,
\end{equation}
which is the generalization of $R\approx -\kappa^2 T_\mu^{\,\,\mu}$ of general relativity. $N^M$ is the matter part and $N_0\equiv \frac{\partial N(X,\bar X)}{\partial X^0} = \bar X^0 \left(\Phi -S^i\frac{\partial \Phi }{\partial S^i}\right)= -3\bar X^0+N_0^{\rm M}$.

\section{New Minimal Supergravity}
The action for pure new minimal supergravity is given in terms of a real linear multiplet $L$ (the chiral $X_0$ doesn't appear in the action) as
\begin{equation}
{\cal L}^{nm} = \left[3L\ln\frac{L}{X^0\bar X^0}\right]_D\,.
 \end{equation}
According to the general Ward identity (2.14) obtained in \cite{Ferrara:2018dyt} , this satisfies
 \begin{equation}
  \overline{{\cal D}}^{\dot \alpha }{\cal E}^L_{\alpha \dot \alpha }= L\,W^L_\alpha\,.
\end{equation}
One can add matter and FI term to this action
\begin{equation}
  {\cal L}^{nm}= \left[ 3L\ln\frac{L}{X^0\bar X^0}\right]_D+ \left[L\,K\right]_D +\xi \left[L\, V\right]_D + \left[W^\alpha W_\alpha \right]_F\,,
\end{equation}
where $K$ is the matter K\"{a}hler potential. The full $A_\mu $ field equation then is split as
\begin{equation}
  {\cal E}^L_\mu+ J_\mu+{\cal E}^W_\mu\approx 0\,.
\end{equation}
Each term satisfies the following Ward identitites
\begin{equation}
  \overline{{\cal D}}^{\dot \alpha }{\cal E}^L_{\alpha \dot \alpha }= L\,W^L_\alpha \,,\qquad  \overline{{\cal D}}^{\dot \alpha }{\cal E}^W_{\alpha \dot \alpha }=-2W_\alpha \,D \,, \qquad
  \overline{{\cal D}}^{\dot \alpha }J_{\alpha \dot \alpha }\approx  L\, W^K _\alpha  \,.
\end{equation}
The relevant field equation that we used here is the field equation of $\chi _\alpha $, the fermionic component of the linear multiplet, which defines $W_\alpha ^{L}$, $W_\alpha ^{K}$ and $W_\alpha ^{\rm nm,FI}=\xi W_\alpha$.

\providecommand{\href}[2]{#2}\begingroup\raggedright\endgroup

\end{document}